\documentclass[12pt,a4paper]{conference}

\usepackage{fancyhdr}
\usepackage{graphicx,amsmath,amssymb,cite}
\usepackage{multind}
\makeindex{author} \makeindex{subject}

\newcommand{\beq}{\begin{equation}}
\newcommand{\eeq}[1]{\label{#1}\end{equation}}
\newcommand{\eeqn}{\end{equation}}

\newcommand{\beqa}{\begin{eqnarray}}
\newcommand{\eeqa}[1]{\label{#1}\end{eqnarray}}
\newcommand{\eeqan}{\end{eqnarray}}

\let\bar=\overbar

\newcommand{\bra}[1]{\left\langle{ #1} \right|}
\newcommand{\ket}[1]{\left| {#1} \right\rangle}

\newcommand{\Dslash}{\not{\hbox{\kern-4pt $D$}}}
\newcommand{\dslash}{\not{\hbox{\kern-2pt $\del$}}}

\newcommand{\msb}{{\bar{\ssstyle M \kern -1pt S}}}

\begin{document}

\Chapter{EXCLUSIVE PRODUCTION OF QUARKONIA IN $pp$ and $p \bar p$
COLLISIONS FAR FROM THE THRESHOLD}
           {Exclusive production of quarkonia}{A. Szczurek \it{et al.}}
\vspace{-6 cm}\includegraphics[width=6 cm]{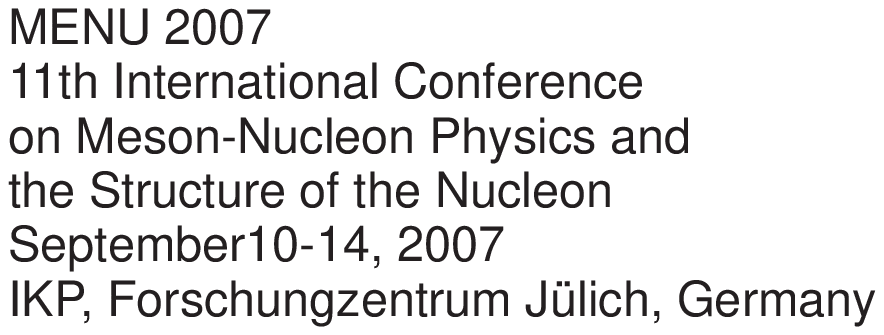}
\vspace{4 cm}

\addcontentsline{toc}{chapter}{{\it A. Szczurek}} \label{authorStart}

\begin{raggedright}

{\it A. Szczurek \footnote{
Institute of Nuclear Physics,
PL-31-342 Krak\'ow, ul. Radzikowskiego 152,\\
Rzesz\'ow University, 
PL-35-959 Rzesz\'ow, ul Rejtana 16A
}}\index{author}{Szczurek, A.}\\
\bigskip\bigskip

\end{raggedright}

\begin{center}
\textbf
{I discuss exclusive production of the $\eta'$, $J/\psi$, 
and $\chi_c(0^+)$ mesons in $pp$ and $p \bar p$ collisions
at high energies. QCD diffractive mechanisms as well as 
nondiffractive mechanisms are discussed. Different 
unintegrated gluon distribution functions (UGDF) are used. 
Some differential distributions are shown and discussed.}
\end{center}

\section{Introduction}

The search for Higgs boson is the primary task for the LHC collider
being now constructed at CERN. Although the predicted cross section
is not small it may not be easy to discover Higgs in inclusive
reaction due to large background in each of the final channel considered.
An alternative way is to search for Higgs
in exclusive or semi-exclusive reactions with large rapidity gaps.
Kaidalov, Khoze, Martin and Ryskin proposed to calculate diffractive
double elastic (both protons survive the collision) production
of Higgs boson in terms of UGDFs
\cite{KMR}. Here I shall present some application of this formalism to
the production of $\eta'$ and $\chi_c(0^+)$ mesons.
More details can be found in Refs.\cite{SPT07,PST07}.

At present, there is ongoing investigations at Tevatron
aiming at measuring the exclusive production of both vector and scalar
quarkonia, but no result is yet publicly available. 

Recently the $J/\psi$ exclusive production in proton-proton and 
proton-antiproton collisions was suggested as a candidate
in searches for odderon exchange \cite{BMSC07}.
In order to identify the odderon exchange one has to consider all
other possible processes leading to the same final channel.
One of such processes, probably dominant, is pomeron-photon or
photon-pomeron fusion \cite{SS07}.

\section{Diffractive production of mesons}
\label{sec:2}


\begin{figure}[!h] 
 \centerline{\includegraphics[width=0.4\textwidth]{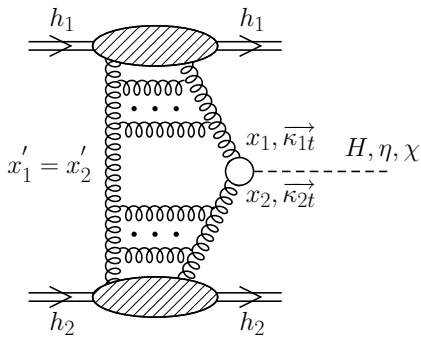}}
   \caption{\label{fig:diffraction_ugdf}
The sketch of the bare QCD mechanism. The kinematical
variables are shown in addition.}
\end{figure}


Following the formalism for the diffractive double-elastic
production of the Higgs boson one can write the amplitude
from Fig.\ref{fig:diffraction_ugdf} as 
\begin{eqnarray}
{\cal M}_{pp \to p M p}^{g^*g^*\to M} =  i \, \pi^2 \int
d^2 k_{0,t} V(k_1, k_2, P_M) \frac{
f^{off}_{g,1}(x_1,x_1',k_{0,t}^2,k_{1,t}^2,t_1)
       f^{off}_{g,2}(x_2,x_2',k_{0,t}^2,k_{2,t}^2,t_2) }
{ k_{0,t}^2\, k_{1,t}^2\, k_{2,t}^2 } \, , 
\label{main_formula}
\end{eqnarray}
where $f's$ are skewed unintegrated gluon distributions. 
For more details see \cite{SPT07}.

\begin{figure}[!h]
\begin{center}
\includegraphics[width=0.5\textwidth]{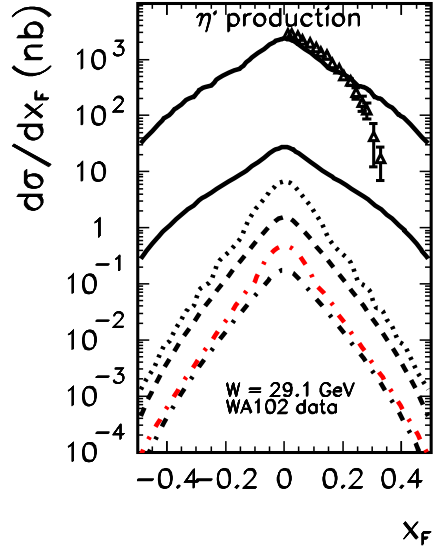}
\end{center}
\caption{$d \sigma / dx_F$ as a function of Feynman $x_F$ for W
= 29.1 GeV and for different UGDFs.
The $\gamma^* \gamma^*$ fusion contribution is shown by
the dash-dotted (red) line (second from the bottom).
The experimental data of the WA102 collaboration \cite{WA102} are shown
for comparison. The dashed line corresponds to the KL distribution, dotted
line to the GBW distribution and the dash-dotted to the BFKL distribution.
The two solid lines correspond to the Gaussian distribution with
details explained in the original paper.
No absorption corrections were included here.
}
\label{fig:dsig_dxF}
\end{figure}

As an example in Fig.~\ref{fig:dsig_dxF} I show the results of
calculations obtained for $M = \eta'$ with several models of UGDF
(for details see \cite{SPT07}) for relatively low energy W = 29.1 GeV.
For comparison I show also the contribution of the
$\gamma^* \gamma^*$ fusion mechanism.
The contribution of the last mechanism is much smaller than
the contribution of the diffractive QCD mechanism.

\begin{figure}[!h]
\begin{center}
\includegraphics[width=0.5\textwidth]{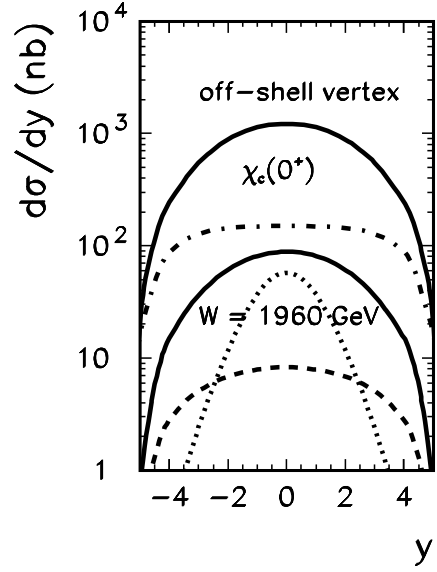}
\end{center}
\caption{$d \sigma / dy$ as a function of rapidity for W
= 1960 GeV and for different UGDFs.
The dashed line corresponds to the KL distribution, dotted
line to the GBW distribution and the dash-dotted to the BFKL distribution.
The two solid lines correspond to the KMR distributions with
details explained in the original paper.
No absorption corrections were included here.
}
\label{fig:dsig_dy_off}
\end{figure}

In Fig.\ref{fig:dsig_dy_off} I show rapidity distribution of scalar
$\chi_c(0^+)$ meson for different UGDFs. Similarly as for the $\eta'$
production a strong dependence on UGDFs can be observed.
\footnote{In Ref.\cite{PST07} we discuss many more uncertainties in theoretical
predictions of exclusive diffractive production.}
A slightly less dependence on UGDFs can be expected for diffractive
Higgs produciton.

\section{Photoproduction of $J/\psi$}


\begin{figure}[!h]    %
\begin{center}
\includegraphics[width=0.3\textwidth]{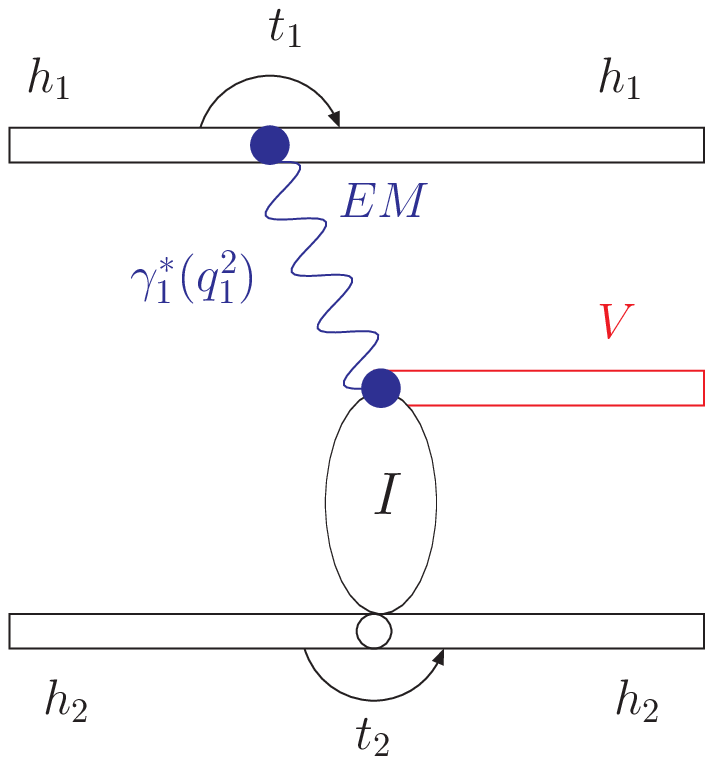}
\includegraphics[width=0.3\textwidth]{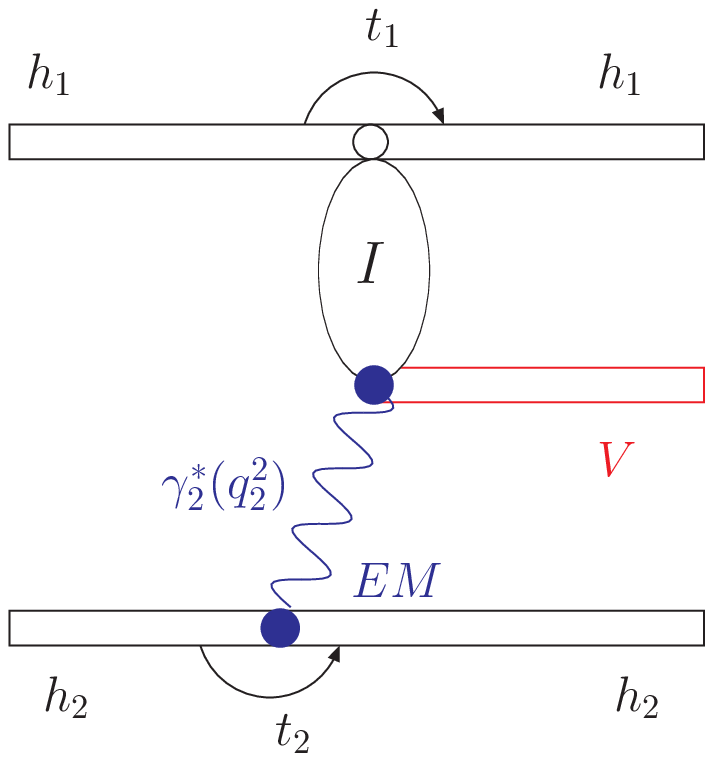}
\end{center}
   \caption{\label{fig:diagram_photon_pomeron}
The sketch of the two mechanisms considered:
photon-pomeron (left) and pomeron-photon (right).
Some kinematical variables are shown in addition.}
\end{figure}


The basic mechanisms leading to the exclusive production
of $J/\psi$ are shown in Fig.\ref{fig:diagram_photon_pomeron}. 
The amplitude for the corresponding $2 \to 3$ process 
can be written as 
\begin{eqnarray}
{\cal M}_{h_1 h_2 \to h_1 h_2 V}^
{\lambda_1 \lambda_2 \to \lambda'_1 \lambda'_2 \lambda_V}(s,s_1,s_2,t_1,t_2) &&= 
{\cal M}_{\gamma IP} + {\cal M}_{IP \gamma} \nonumber \\
&&= \bra{p_1', \lambda_1'} J_\mu \ket{p_1, \lambda_1} 
\epsilon_{\mu}^*(q_1,\lambda_V) {\sqrt{ 4 \pi \alpha_{em}} \over t_1}
{\cal M}_{\gamma^* h_2 \to V h_2}^{\lambda_{\gamma^*} \lambda_2 \to \lambda_V \lambda_2}
(s_2,t_2,Q_1^2)   \nonumber \\
&& + \bra{p_2', \lambda_2'} J_\mu \ket{p_2, \lambda_2} 
\epsilon_{\mu}^*(q_2,\lambda_V)  {\sqrt{ 4 \pi \alpha_{em}} \over t_2}
{\cal M}_{\gamma^* h_1 \to V h_1}^{\lambda_{\gamma^*} \lambda_1 \to \lambda_V \lambda_1}
(s_1,t_1,Q_2^2)  \, . \nonumber \\
\label{Two_to_Three}
\end{eqnarray}
%
%
%
The amplitude of the $\gamma^* p \to J/\psi p$ was parametrized
to describe data measured recently at HERA \cite{ZEUS_JPsi,H1_JPsi}.

The differential cross section is calculated as
\begin{equation}
d \sigma = { 1 \over 512 \pi^4 s^2 } | {\cal M} |^2 \, dy dt_1 dt_2
d\phi \, ,
\end{equation}
where $y$ is the rapidity of the vector meson, and $\phi$ is the angle
between outgoing protons.
Notice that the interference between the two mechanisms $\gamma IP$
and $IP \gamma$ is proportional to $e_1 e_2$ 
and introduces a charge asymmetry.

\begin{figure}[!h]   
 \centerline{\includegraphics[width=0.5\textwidth]{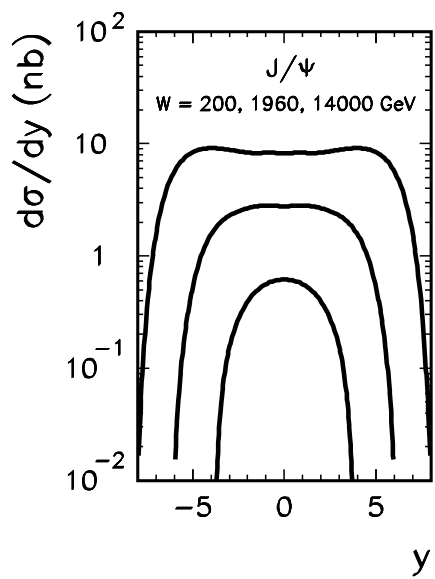}}
   \caption{ \label{fig:dsig_dy_energy}
$d \sigma / dy$ for exclusive $J/\psi$ production
as a function of $y$ for RHIC, Tevatron and LHC energies.
No absorption corrections were included here.}
\end{figure}

In Fig.\ref{fig:dsig_dy_energy} I collect rapidity distributions
for different energies relevant at RHIC, Tevatron and LHC. One observes an
occurence of a small dip in the distribution at midrapidities at LHC energy.
One should remember, however, that the distribution for the LHC energy is
long-distance extrapolation of the $\gamma^*p \to J/\psi p$
amplitude (or cross section) to unexplored yet experimentally energies.
Therefore a real experiment at Tevatron and LHC would help to constrain 
cross sections for $\gamma p \to J/\psi p$ process.


\begin{figure}[!h]    %
\begin{center}
\includegraphics[width=0.3\textwidth]{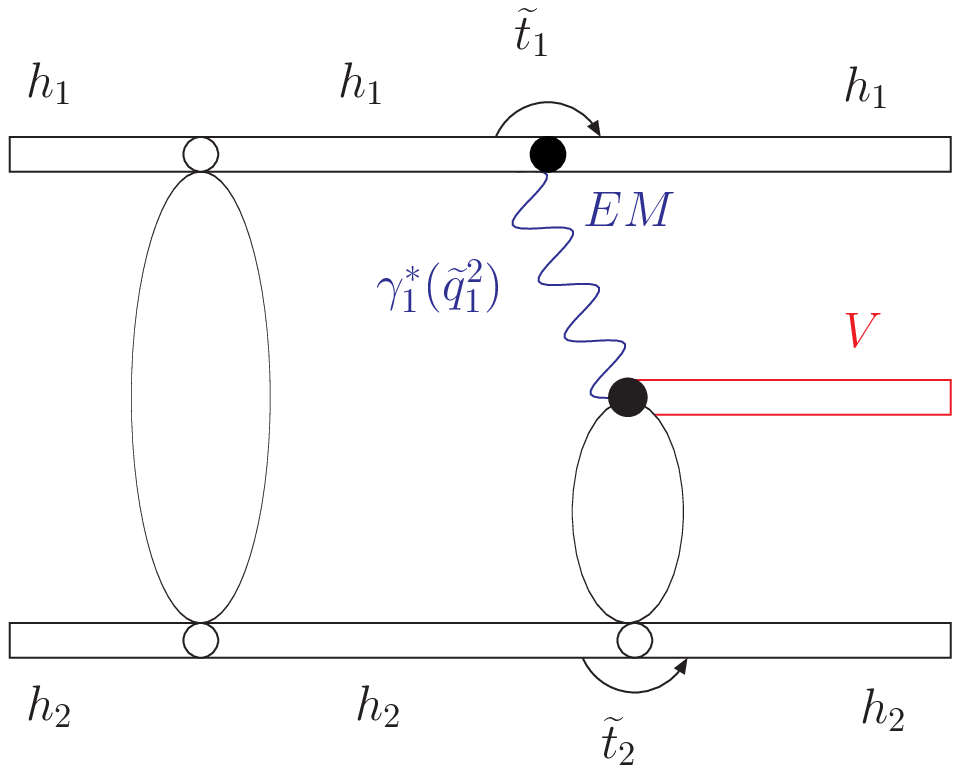}
\includegraphics[width=0.3\textwidth]{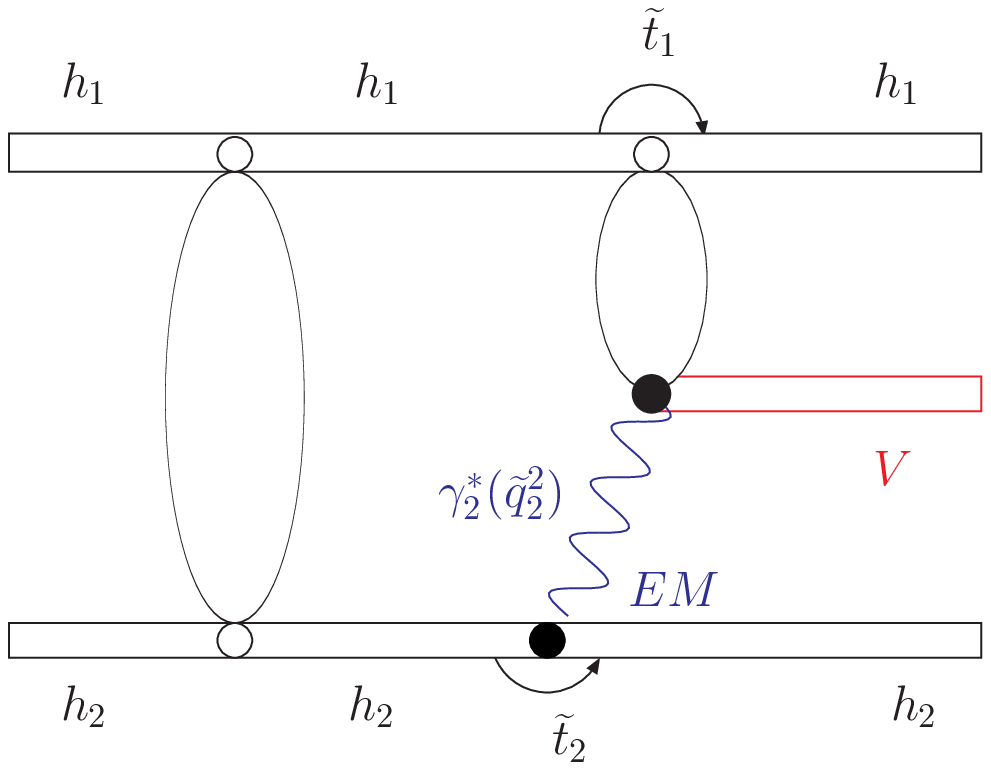}
\end{center}
   \caption{\label{fig:diagram_rescattering}
The sketch of the elastic rescattering amplitudes.
Some kinematical variables are shown in addition.}
\end{figure}

The parametrization of the $\gamma^* p \to V p$ amplitude which
describes corresponding experimental data (see \cite{SS07}) includes
effectively absorption effects due to final state $Vp$ interactions.
In the $p p \to p p J/\psi$ ($p \bar p \to p \bar p J/\psi$)
reaction the situation is more complicated as here $pp$ (or $p \bar p$)
strong rescatterings occur in addition. In Ref.\cite{SS07} we have
included only elastic rescatterings shown schematically in 
Fig.\ref{fig:diagram_rescattering}.


\section{Summary}
\label{sec:5}

In contrast to diffractive Higgs production, in the case of meson
production the main contribution to the diffractive amplitude comes
from the region of very small gluon transverse momenta and very small
longitudinal momentum fractions. In this case application of
Khoze-Martin-Ryskin UGDFs seems not justified and we have to rely
on UGDFs constructed for this region.

The existing models of UGDFs predict cross section much smaller
than the one obtained by the WA102 collaboration at the
center-of-mass energy W = 29.1 GeV. This may signal presence of
subleading reggeons at the energy of the WA102 experiment or
suggest a modificaction of UGDFs in the nonperturbative region of
very small transverse momenta.

%
The $\gamma^* \gamma^*$ fusion may be of some importance only at
extremely small four-momentum transfers squared for the $\eta'$
production and is practically negligible for the $\chi_c(0^+)$ production.

It was shown in \cite{SS07} that at the Tevatron energy one can study
the exclusive production of $J/\psi$ at the photon-proton center-of-mass
energies 70 GeV $ < W_{\gamma p} < $ 1500 GeV, i.e. in the unmeasured region
of energies, much larger than at HERA. At LHC this would be correspondingly
200 GeV $ < W_{\gamma p} < $ 8000 GeV. At very forward/backward
rapidities this is an order of magnitude more than possible
with presently available machines.

An interesting azimuthal-angle correlation pattern has been obtained
due to the interference of photon-pomeron and pomeron-photon 
helicity-preserving terms.

We have estimated also absorption effects. In some selected configurations
the absorption effects may lead to the occurence of diffractive minima.
The exact occurence of diffractive minima depends on the values
of the model parameters. Such minima are washed out when integrated
over the phase space or even its part. 


\bigskip
\section{Acknowledgements}
I thank organizers of MENU2007 for very efficient organization
of the conference and hospitality.
The collaboration with Roman Pasechnik, Oleg Teryaev and Wolfgang Sch\"afer
on the issues presented here is acknowledged.
A partial support by  the MEiN research grant~
1~P03B~028~28 (2005-08).




\begin{thebibliography}{100}



\bibitem{KMR}
V.A. Khoze, A.D. Martin and M.G. Ryskin, Phys. Lett. B {\bf 401},
330 (1997);\\
V.A. Khoze, A.D. Martin and M.G. Ryskin, Eur. Phys.
J. C {\bf 23}, 311 (2002);\\
A.B. Kaidalov, V.A. Khoze, A.D. Martin
and M.G. Ryskin, Eur.\ Phys.\ J.\ C {\bf 31}, 387 (2003)
[arXiv:hep-ph/0307064];\\
A.B. Kaidalov, V.A. Khoze, A.D. Martin and M.G. Ryskin,
Eur. Phys. J. C {\bf 33}, 261 (2004);\\
V.A. Khoze, A.D. Martin, M.G. Ryskin and W.J. Stirling,
Eur. Phys. J. C {\bf 35}, 211 (2004).

\bibitem{SPT07}
  A.~Szczurek, R.~S.~Pasechnik and O.~V.~Teryaev,
  Phys.\ Rev.\  D {\bf 75}, 054021 (2007)
  [arXiv:hep-ph/0608302].

\bibitem{PST07}
R.~S.~Pasechnik, A.~Szczurek and O.~V.~Teryaev,
  [arXiv:0709.0857].

\bibitem{BMSC07}
A. Bzdak, L. Motyka, L. Szymanowski and J.-R. Cudell, arXiv:
hep-ph/0702134.

\bibitem{SS07}
W. Sch\"afer and A. Szczurek, arXiv:0705.2887, 
Phys. Rev. {\bf D76}, 094014.

\bibitem{WA102}
D. Barberis et al. (WA102 collaboration),
Phys. Lett. {\bf B422}, 399 (1998).

\bibitem{ZEUS_JPsi}
  S.~Chekanov {\it et al.}  [ZEUS Collaboration],
  Eur.\ Phys.\ J.\  C {\bf 24}, 345 (2002).

\bibitem{H1_JPsi}
  A.~Aktas {\it et al.}  [H1 Collaboration],
  Eur.\ Phys.\ J.\  C {\bf 46}, 585 (2006).


\end{thebibliography}
\end{document}